\documentclass[journal = nalefd, manuscript = letter]{achemso}
\usepackage{graphicx,epsf,epsfig}
\usepackage{dcolumn}
\usepackage{bm}
\usepackage{color}
\usepackage{ulem}
\usepackage{subfigure}
\usepackage{upgreek}

\title[\texttt{achemso}]
{Efficient Exciton Transport in Strongly Quantum-Confined Silicon Quantum Dots}

\author{Zhibin Lin}
\email{zlin@mines.edu}
\affiliation{Department of Physics, Colorado School of Mines, Golden, CO 80401, USA}
\alsoaffiliation{National Renewable Energy Laboratory, Golden, Colorado 80401, USA}
\author{Huashan Li}
\email{huali@mines.edu}
\affiliation{Department of Physics, Colorado School of Mines, Golden, CO 80401, USA}
\author{Alberto Franceschetti}
\email{Alberto.Franceschetti@nrel.gov}
\affiliation{National Renewable Energy Laboratory, Golden, Colorado 80401, USA}
\author{Mark T. Lusk}
\email{mlusk@mines.edu}
\affiliation{Department of Physics, Colorado School of Mines, Golden, CO 80401, USA}

\begin{document}

% Abstract%
\begin{abstract}
First-order perturbation theory and many-body Green function analysis are used to quantify the influence of size, surface reconstruction and surface treatment on exciton transport between small silicon quantum dots. Competing radiative processes are also considered in order to determine how exciton transport efficiency is influenced. The analysis shows that quantum confinement causes small ($\sim$1 nm) Si quantum dots to exhibit exciton transport efficiencies far exceeding that of their larger counterparts. We also find that surface reconstruction significantly influences the absorption cross section and leads to a large reduction in both transport rate and efficiency. Exciton transport efficiency is higher for hydrogen passivated dots as compared with those terminated with more electronegative ligands. This is because such ligands delocalize electron wave functions towards the surface and result in a lower dipole moment. Such behavior is not predicted by F\"orster theory, built on a dipole-dipole approximation, because higher order multi-poles play a significant role in the exciton dynamics.

\end{abstract}
\maketitle


%Introduction
%%%%%%%%%%%%%%%%%%%%%
Recent advances in the synthesis of semiconductor quantum dots (QDs) have opened up an intriguing opportunity for exploiting quantum confinement to inexpensively improve photovoltaic energy conversion efficiency \cite{Nozik_PE_2002, Jurbergs-APL-2006, Beard_NL_2007, Timmerman-NP-2008, Nozik-NL-2011, Lin-ACSNano-2011}.  With tunable optical gaps and strong absorption cross-sections, the manufacture-friendly assemblies composed of QDs offer a design panacea for efficient exciton generation, but energy transfer out of such materials is still problematic. Charge separation paradigms are being explored in which polymeric charge transport networks are interspersed with the QDs, but these are currently limited by the low efficiency of charge separation and carrier transport inherent with the polymers~\cite{Yu-JAP-1995,Dittmer-AM-2000}.  Recently though, Lu et al.  \cite{Lu-NL-2007, Lu-NL-2009} proposed an alternative paradigm in which nonradiative energy transfers occur from dot to dot and ultimately to high mobility carrier transport channels. This promising approach depends critically on the efficiency of such exciton transport as compared with available relaxation mechanisms. This has motivated us quantify the ways in which exciton transport efficiency can be optimized by varying the size, spacing and surface termination of QDs.

We chose to study silicon-based QDs because they are environmentally benign, not resource challenged, and have received a great deal of attention lately for photovoltaic and light-emitting diode applications \cite{Lu-Nature-1995, Cheng-NL-2010, Puzzo-NL-2011}. For instance, carrier multiplication through multiple-exciton generation (MEG)  has been recently demonstrated in colloidal silicon QDs  \cite{Beard_NL_2007} and for Si QDs embedded in a SiO2 matrix \cite{Timmerman-NP-2008,Timmerman-NNano-2011}. This promise, though, has been somewhat shadowed by theoretical predictions~\cite{Delerue-PRB-2007} that resonant energy transfer between large ($\sim$2-4 nm) Si QDs is possible only when they are almost in contact because of the relative efficiency of radiative relaxation. Recent work on MEG in QDs \cite{Beard_NL_2010, Lin-ACSNano-2011} suggests that there is a thermodynamic advantage to working with small QDs, and these are more strongly influenced by terminating groups as well. We have therefore focused on small Si QDs with diameters of $\sim$1.0--1.5 nm.

Provided that the QDs are sufficiently well separated, exciton hopping can be considered within the framework of F\"orster resonance energy transfer (FRET), wherein Coulomb interactions are approximated using either dipole-dipole \cite{Froster-AP-1948} or higher-order multipole expansions \cite{Dexter-JCP-1953}.  However, it is not clear that such approximations are appropriate in the case of semiconductor QDs since F\"orster theory (FT) assumes a point dipole interaction while QDs have a finite size. When nanometer-scale ligands are all that separate dots, the applicability of FT is particularly questionable. 

Several recent theoretical studies have aimed to examine the validity of FT in semiconductor QDs. Curutchet et al. \cite{Scholes-JPCC-2008} examined the electronic coupling between two 3.9 nm CdSe QDs and also between a QD and a chlorophyll molecule. They found that the dipole approximation works rather well for spherical QDs even at contact separations.  Schrier and Wang \cite{Wang-JPCC-2008} studied the shape dependence of resonant energy transfer between CdSe QDs and found that the dipole-dipole interaction underestimates the coupling between linearly oriented nanorods and overestimates the coupling between parallel nanorods. Roi and Rabani \cite{Rabani-JCP-2008} investigated the relative contribution of various multipole interactions to FRET.  Allan and Delerue \cite{Delerue-PRB-2007} studied the resonant energy transfer in InAs and Si QDs using a tight-binding approach and found that, for direct-gap InAs, the transfer rate is well described by FT, whereas for Si QDs FT fails at small dot separations due to the weakness of dipolar transitions, a reflection of the indirect nature of Si. This has motivated us to consider a many-body Green function approach that is not based on the assumption that dipole-dipole interactions dominate, a central tenant of FT.

% Approach
%%%%%%%%%%%%%%%%%%%%%%%%

All Si QD geometries were optimized within a local density approximation using standard density functional theory (DFT). The influence on exciton dynamics of five terminating ligands was considered: H, CH$_{3}$, F, Cl and OH. A many-body Green function approach, explicitly accounting for the quasi-particle effects and electron-hole interactions, was then employed to calculate the excitonic eigenstates in these quantum-confined structures. In particular, the electron self-energy was calculated within the GW$_{{\it f}}$ approximation \cite{Tiago-PRB-2006,Louie-PRB-1986} and the excitonic characteristics were accounted for by solving the Bethe-Salpeter equation (BSE) \cite{Onida-RMP-2002, Louie-PRB-2000, Tiago-PRB-2006}. The lowest excitonic energies obtained in our calculations (Table 1) agree well with earlier GW-BSE calculations\cite{Tiago-PRB-2006}. Fig.\ 1 illustrates the exciton transport process between two identical Si QDs considered in this work. Within first-order perturbation theory, the exciton transport rates are quantified using Fermi's Golden Rule:

%Equation for Fermi's golden rule%
%
\begin{equation}
 \label{eq:FGR}
\Gamma = \frac{2\pi}{\hbar}\sum_{i}p(i)\vert\langle \Phi_{i} \vert W\vert \Phi_{f}\rangle\vert^{2}\delta(E_{i}-E_{f})
\label{FGR}
\end{equation}
where $\Phi_{i}$ and $\Phi_{f}$ are the many-body electron wave functions for initial ({\it i}) and final ({\it f}) excitonic states. The term, {\it p(i)}, accounts for the room-temperature Boltzmann occupation of the initial state. $E_{i}$ and $E_{f}$ are the energies of the initial and final excitonic states. Energy conservation during the exciton transport process is enforced by replacing the delta function of \ref{FGR} with a Gaussian profile that has a variance of 10 meV. This accounts for the thermal broadening effects of excitonic levels due to electron-phonon coupling in Si QDs \cite{Delerue-PRB-2007, Delerue-Nanostructures-2004}. \ref{FGR} takes into account both direct and exchange Coulomb interaction between the initial and final states as well as the screening of the Coulomb interaction due to other excitonic states. A complete account of screening was not undertaken because of the computational demands associated with the mixing of the large number of excitonic states. Instead, we used a factor \cite{Delerue-PRB-2007, Rabani-JCP-2008}, $\eta$, as a means of approximating the screening of Coulomb interactions between initial and final states. This factor was computed from the effective dielectric constant, $\epsilon$, defined as the ratio of unscreened and screened Coulomb energies of the lowest triplet excitonic state from GW-BSE calculations. The effective dielectric constants for all Si dots considered in this work are found to be small, less than 2 as compared to a value of 11.4 for bulk silicon \cite{Faulkner-PR-1969}. This is consistent with previous calculations.~\cite{Chelikowsky-PRL-2003} for Si QDs of similar sizes. The effective dielectric constants for all hydrogen terminated Si QDs are provided in Table 1.

The Coulomb integral of Eq. 1 can be approximated using a modified multipole expansion in which a local field factor is applied to each term as an approximate means of extending the validity of the expansion down to the length scales of interest in the current setting.~\cite{Rabani-JCP-2008} The fill factor is
\begin{equation}
 f_{l}(x) = \frac{2l+1}{(x+1)l +1} 
 \label{eq:FillFactor}
\end{equation}
Here $l$ is the order of the pole,  x=$\epsilon_{QD}$/$\epsilon_{0}$, and $\epsilon_{QD}$ and $\epsilon_{0}$ are the dielectric constants of the QD and  medium, respectively. We assume a vacuum outside of the QDs--i.e. $\epsilon_{0}$ = 1 so that, for example, the local field factor of a dipole is $f_{1}(\epsilon_{QD})=3/(\epsilon_{QD}+2)$.  Eq. 1 can be then re-written as
\begin{equation}
 \label{eq:FOR}
\Gamma_{ET} = \frac{2\pi}{\hbar} \eta^{2} \sum_{i}p(i)\vert\sum_{j,k}B^{i}_{j}B^{f}_{k}(K_{c^{A}_{j},v^{A}_{j},v^{B}_{k},c^{B}_{k}}-K_{c^{A}_{j},c^{B}_{k},v^{B}_{k},v^{A}_{j}})\vert^{2}\delta(E_{i}-E_{f})
 \label{FOR}
\end{equation}
In this equation, $B^{i}_{j}$ and $B^{f}_{k}$ are the BSE eigenvectors for the transition in which the initial state is exciton {\it j} (single-particle valence orbital $\psi^{v}_{j}$ and single-particle conduction orbital $\psi^{c}_{j}$ on dot A) and the final state is exciton {\it k} ($\psi^{v}_{k}$ and $\psi^{c}_{k}$ on dot B)~\cite{Tiago-PRB-2006}. Due to weak dielectric screening of the small Si QDs considered in this work (Table I), the parameter $\eta$ is taken as $f^{2}_{1}(\epsilon_{QD})$ corresponding to the case of two identical QDs under the dipole-dipole approximation. The first and second term appearing inside the parentheses of \ref{FGR} are the bare direct and exchange Coulomb integrals. We find that the exchange Coulomb interactions have negligible contributions to the exciton transport for the range of dot separations considered here. The Coulomb integrals are defined as~\cite{Tiago-PRB-2006} 
\begin{equation}
 \label{eq:K}
K_{m,n,p,q} =\sum_{\sigma,\sigma\prime}\int\int {\rm d}{\bf r} {\rm d} {\bf r}' \ \psi^{*}_{m}({\bf r},{\bf \sigma}) \psi_{n}({\bf r},{\bf \sigma}) \frac{e^2}{\vert {\bf r} - {\bf r}'\vert }\psi^{*}_{p}({\bf r}',{\bf \sigma}') \psi_{q}({\bf r}',{\bf \sigma}'),
 \label{K}
\end{equation}

 \begin{table}
 \caption{\label{tab4} Lowest singlet excitonic energy, $E^{X}_{1}$, photoluminescence lifetime, $\tau_{PL}$, and static dielectric constant, $\epsilon_{QD}$, for four hydrogen-passivated Si QDs.}
 \begin{tabular}{|c|c|c|c|c|c|c|}\hline
 QD&Diameter, nm&$E^{X}_1$, eV&$\tau_{PL}$, s&$\epsilon_{QD}$\\ \hline
Si$_{17}$H$_{36}$&0.9 & 4.8 & 3.0$\times 10^{-7}$ & 1.12 \\
Si$_{35}$H$_{36}$&1.1 & 4.1 & 8.5$\times 10^{-7}$ & 1.15 \\
Si$_{66}$H$_{64}$&1.3 & 3.7 & 4.1$\times 10^{-6}$ & 1.20 \\
Si$_{66}$H$_{40}$&1.3 & 2.5 & 1.5$\times 10^{-3}$ & 1.12 \\
Si$_{87}$H$_{76}$&1.5 & 3.2 & 6.9$\times 10^{-5}$ & 1.27 \\ \hline
 \end{tabular}
 \end{table}

Fig. 2 (a) shows the exciton transport rates in four hydrogen-passivated Si QDs for a range of center-to-center separations. Four QD sizes were considered: 0.9 nm (Si$_{17}$H$_{36}$), 1.1 nm (Si$_{35}$H$_{36}$), 1.3 nm (Si$_{66}$H$_{64}$), and 1.5 nm (Si$_{87}$H$_{76}$). The results show that, for a wide rage of fixed center-to-center separation, the rate of exciton transport increases as dot size decreases. Moreover, these rates increase rapidly as the center-to-center distance is decreased. Since the exciton mobility scales as the ratio of square of hopping distance per hopping time, this better than logarithmic increase implies that the highest mobility is achieved by making dots as small and as close together as possible. The increase in transport rate with decreasing dot size is due to the effect of quantum confinement on the intra-dot oscillator strengths. As shown by Dexter \cite{Dexter-JCP-1953}, within dipole-dipole approximation the direct Coulomb interactions in \ref{FOR} could be associated with the products of intra-dot oscillator strengths of initial and final excitons. As the dot size decreases, the oscillator strength in small Si QDs increases drastically\cite{Trani-PRB-2005, Ramos-PSSb-2005}, leading to the strong enhancement in the exciton transport rates. 

The results of our calculations also show that FT breaks down for dots separated by less than 2 nm (Fig.\ 2(a)). The deviations is due to the influence of higher-order multipole interactions which are missing from FT. At large distances, though, the calculated exciton transport rates are well described by the $1/R^{6}$ scaling of FT quantified using the results obtained from our GW-BSE calculations. Even this region of correspondence may seem surprising since the dipole interactions underlying FT are zero for indirect gap materials as a result of momentum conservation.\cite{Delerue-PRB-2007} However, quantum confinement is very significant for the small QDs considered here, and materials thus constrained have a pseudo-direct dipolar transition\cite{Hybertsen-PRL-1994,Kovalev-PRL-1998,Ramos-PSSb-2005, Klimov-PRL-2008}. In addition to this observation, it is noted that at short distance of 2 nm, the smallest 0.9 nm dot has slightly smaller transport rate than 1.1 nm dot, opposite to the trend predicted from higher-order multipole interactions. This suggests that the specific orientation of the small dot becomes important at close dot separations and the assumption of spherical geometry of the dot in FT at these distances might fail.

While the above trends in transport rate are significant on their own, they do not necessarily translate into gains in transport efficiency. In particular,  photoluminescence rate is also influenced by QD size, separation, and terminating structure. In order to quantify how transport rates change relative to the rate of exciton recombination, we define the {\it exciton transport efficiency}, $p=\Gamma_{ET}/(\Gamma_{ET}+\Gamma_{PL})$, where $\Gamma_{PL}$ is the room-temperature photoluminescence (PL) rate defined as $1/\tau_{PL}$. $\tau_{PL}$ is the average PL lifetime calculated from the oscillator strengths from the excitonic states \cite{Dexter-SSP-1958,Califano-PRB-2007}:
\begin{equation}
 \label{eq:PL}
\Gamma_{PL} =\frac{1}{\tau_{PL}}= \sum_{i}p(i)\frac{4e^{2}\eta E^{3}_{i}}{3c^{3}\hbar^{3}}\vert M_{i} \vert^{2}
\label{FGR2}
\end{equation}
where $E_{i}$ is the excitonic energy of excited state $i$, $\eta$ is equal to $f^{2}_{1}(\epsilon_{QD})$ as defined earlier, and $M_{i}$ is the dipole matrix element between excitonic state $i$ and ground state. . 

This measure of transport efficiency is provided in Fig.\ 2(b) for four hydrogen-passivated Si QDs. For dot separations of less than 3 nm, the transport efficiencies for all four dot sizes are nearly identical and are greater than 0.95.  This makes sense since the transport rates at these distances are much larger, sometimes several orders of magnitude, than the associated PL rates. Within this regime, for instance, it is more probable for 13 exciton jumps to occur than for the exciton to recombine. The transport efficiency between the smaller dots, though, remains greater than 50\% out to separations of 8 nm--i.e. at more than six times the dot diameters. Considered within the dipole-dipole approximation, exciton transport rate correlates with the overlap between the emission spectra and the absorption cross section \cite{Delerue-PRB-2007}.   The emission component appears in both the transport and PL rates and so falls out of the efficiency ratio. The efficiency is therefore determined solely by the absorption cross section at the emission energy $\hbar\omega$, thus could be related to the average PL rate which increases as the dot size decreases due to quantum confinement effects (Table I). This explains why transport efficiency increases as dot size decreases. In addition, the transport efficiency appears to exhibit an asymptotic behavior towards the dots around 1 nm. This is related to the fact that the change in the oscillator strengths due to reduction of dot size is more significant for large dots, for instance it is an order of magnitude from 1.5 nm to 1.3 nm while only by a factor of 2 from 1.1 nm to 0.9 nm. Somewhat at odds with this trend, though, the 1.1 nm Si QD actually has a slightly higher transport efficiency than the 0.9 nm dot. This is because the efficiency is also affected by the emission energy, $\hbar\omega$. Within the dipole-dipole approximation, the ratio $\Gamma_{PL}/\Gamma_{ET} $ is proportional to $R^{6}\omega^{4}/\sigma(\hbar\omega)$~\cite{Delerue-PRB-2007}. Both the absorption cross section, $\sigma(\hbar\omega)$, and the emission energy, $\hbar\omega$, for the 0.9 nm dot are larger than the corresponding terms in 1.1 nm dot due to the quantum confinement effects. At the same dot separation, $R$, the ratio $\Gamma_{PL}/\Gamma_{ET} $ is slightly larger for the 0.9 nm dot, resulting in a slightly smaller transport efficiency. This effect is more visible at large dot separations as shown the figure. Note that the size trend reversal between transport rate and transport efficiency is only exhibited because the two dots in question are near the asymptotic, small dot limit. Overall, we find that the exciton transport efficiency increases as dot size decreases. This suggests that assemblies composed of smaller dots could result in higher quantum yields for the electric current collected per photon absorbed.

Si QDs may undergo surface reconstruction, and this influences exciton transport efficiency. In particular, 2$\times$1-like surface reconstruction can occur on the (100) facets of Si QDs reducing the number of surface dangling bonds. To analyze the impact of such reconstructions, we compared the transport efficiency of reconstructed Si$_{66}$H$_{40}$ dots with their unreconstructed Si$_{66}$H$_{64}$ counterparts. Fig. 3 (a, b) shows that surface reconstruction leads to a reduction in both the exciton transport rate and efficiency. This is due to the fact that reconstruction reduces the absorption cross-section, quantified in Fig. 3(c) by GW-BSE calculations. The reduction is due to the delocalization of electronic wave functions in response to surface strain generated by surface reconstruction. This behavior is similar to increasing the QD size shown in Fig. 2 as both the surface reconstruction and an increase in dot size decrease the oscillator strengths of the lowest excitonic states. These excitonic states are heavily involved in the exciton transport due to large probability of occupying these states at room temperature according to the Boltzmann distribution.

The influence of surface treatment on exciton transport rate was investigated using Si$_{35}$ QDs passivated with five different ligands: H, CH$_{3}$, F, Cl and OH. The results are summarized in Fig. 4. Earlier calculations for OH passivation using DFT with a B3LYP exchange/correlation functional found that the HOMO-LUMO transition is dipole forbidden\cite{Zhou-NL-2003}. Our calculations, which explicitly take many-body and excitonic effects into account, show that the mixing of other dipole-allowed transitions makes the lowest excitonic state weakly allowed--i.e. that there is a non-zero oscillator strength. The significant drop in  exciton transport rate that results from surface passivant of OH instead of H  (Fig. 4), is directly related to much weaker oscillator strengths for the lowest excitonic states in OH-passivated Si QDs. The influence of the more electronegative OH passivants on the surface leads to delocalization of electron wave functions to the surface region. In particular, the symmetry of HOMO level is found to change from T2 to T1 after the substitution OH, creating a forbidden transition between HOMO and LUMO (A1). The effect of OH passivation on the oscillator strengths for optical transitions might be related to previous experimental observations of a long, 10$^{-3}$ s, photoluminescence lifetime for 1-to-2 nm Si QDs with an oxide shell. \cite{Brus-Science-1993, Brus-JACS-1995} The symmetry change in the HOMO orbital also occurs with three other terminating groups--i.e. CH$_{3}$, F, and Cl in this work. As a result, the dipole transitions between HOMO and LUMO orbitals for these QDs are very weak, leading to much smaller exciton transport rates shown in the top panel of Fig. 4. The impact of ligand substitution on the exciton transport is further investigated by varying the surface coverage of the OH passivants on the Si QD (Supporting Information). The distortion of the associated wave functions, due to ligand substitution, is shown to be the cause of the significant changes to the exciton transport rate and efficiency.

Fig. 4(a) further indicates that the exciton transport rates for F and OH passivated Si QDs exhibit a strong $1/R^{14}$ dependence, and this is also true for CH$_{3}$ and Cl passivation at small dot separation (less than 4 nm). This scaling relationship suggests that for F and OH passivated Si QDs the leading term in the multipole expansion of the Coulomb interaction is an octupole-octupole interaction; quadrupole interactions vanish due to the Td symmetry of the QD structure. Nevertheless the octupole-octupole interaction still dominates the range of dot separation considered in this study, and FT gives an incorrect prediction of exciton transport rates at all distances. The same conclusions apply to dots passivated with CH$_{3}$ and Cl in which the $1/R^{6}$ scaling of FT begins to take over only at dot separations of approximately 4 and 9 nm, repspectively. For CH$_{3}$ passivation, even though the lowest excitonic state has a very weak oscillator strength due to dipole-forbidden HOMO-LUMO transition, the next excitonic states (only $\sim$180 meV higher) have oscillator strengths that are roughly six orders of magnitude higher. As a result, the exciton transport rates have trends as predicted from a dipole-dipole approximation beyond 4 nm (Fig. 4(a)) where the octupole-octupole interactions become weak.

We now turn our attention to the influence of surface termination on transport efficiency. The analysis is summarized in Fig. 4(b). The efficiency is nearly unity for all five passivations when the dots are separated by less than 3 nm. At larger distances, though, the H-passivated dots exhibit a much higher transport efficiency. Dots passivated with more electronegative F, Cl and OH give rise to longer averaged PL lifetimes, potentially increasing the time window for excitons to be transported. However, the ratios of $\Gamma_{PL}/\Gamma_{ET}$ still remain much larger than that of H passivation. This leads to less efficient exciton transport for these passivants. Nevertheless, for F and OH passivated Si QDs, long PL lifetimes in the lowest exciton states due to weak dipole moments still result in a relatively high transport efficiency at dot separation less than 5 nm. For CH$_{3}$, on the other hand, the lowest excitonic energy is about 4.2 eV, very close to that of H passivated Si$_{35}$. As discussed for the size effect on the exciton transport efficiency, the efficiency can be related to the averaged PL lifetime within the dipole-dipole approximation. Therefore the low transport efficiency in CH$_{3}$ passivated Si$_{35}$ can be explained by its long PL lifetime, 2$\times 10^{-3}$ s, as compared with 8.5$\times 10^{-7}$ s in H passivated Si$_{35}$. 

In summary, we have investigated the exciton transport in $\sim$1--1.5 nm Si nanocrystal QDs using {\it ab initio} many-body approach. Within the first-order perturbation theory, we found that small hydrogen terminated Si QDs exhibit a very high efficiency of the exciton transport as a result of strong quantum confinement. We also show that the surface reconstruction significantly reduces the exciton transport efficiency.  Our findings suggest that small Si QDs of size about 1 nm could speed up the exciton transport process, potentially facilitating the photoconversion process in Si QD assemblies for an improved photovoltaic efficiency. In particular, we find that the exciton transport could be very efficient for hydrogen terminated Si QDs of size less than 1.5 nm in diameter. The transport distance for these small Si QDs could be much larger than the typical dot size, e.g. the transport efficiency still exceeds 0.5 at 8 nm. We also examined several other surface terminations, including CH$_{3}$, F, Cl, OH, in Si QDs and found that, at the same QD separation, exciton transport is more efficient in H-terminated Si QDs than QDs passivated with the other types of surface ligands. 

Interestingly, for hydrogen terminated Si QDs the F\"orster's $1/R^{6}$ expression agrees very well with exciton transport rates calculated from first principles except at short distances ($\sim$2 nm) where multipole interactions becomes important. On the other hand, we find that for other surface passivations (F, Cl and OH) the F\"orster's theory fails to predict correct exciton transport rates due to the vanishing dipole moments in these structures. Exciton transport in F- and OH-passivated Si QDs is dominated by octupole-octupole interaction, and associated long PL lifetimes in the lowest exciton states are due to weak dipole moments. This results in a relatively high transport efficiency at dot separations of less than 5 nm.

\begin{acknowledgement}
This work was supported by the Renewable Energy Materials Research Science and Engineering Center (NSF Grant No. DMR-0820518) at the Colorado School of Mines and the National Renewable Energy Laboratory. The calculations were carried out using the high performance computing resources provided by the Golden Energy Computing Organization at the Colorado School of Mines (NSF Grant No. CNS-0722415). 
\end{acknowledgement}

\begin{suppinfo}
The Supporting Information Section includes the discussion on the influence of surface coverage of OH ligands on the exciton transport in Si QDs.
\end{suppinfo}

\bibliographystyle{achemso}

\newpage

\section*{Figure captions}
% Add figure captions
\begin{description}

\item[Figure \ 1]
Schematic diagram of an exciton, initially located on dot A, hopping to neighboring dot B with corresponding energy levels shown in the bottom panel. The ground state for a single QD is denoted by 0 in a Si QD with 1, 2, and n lowest, second lowest and {\it n}th excitonic states. Degenerate excitonic states may exist and are fully taken into account in the calculation of transport rates. 

\item[Figure \ 2]
Exciton transport rate and efficiency as a function of dot separation in three hydrogen passivated Si QDs: 1.1 nm (Si$_{35}$H$_{36}$), 1.3 nm (Si$_{66}$H$_{64}$), and 1.5 nm (Si$_{87}$H$_{76}$). Solid curves in top panel (a) are obtained directly from F\"orster theory using the dipole-dipole approximation wherein transport rates scale as $1/R^{6}$. Plot in bottom panel (b) is of exciton transport efficiency, defined as the ratio of exciton transport rate and total exciton decay rate.

\item[Figure \ 3]
(a) Exciton transport rate, (b) efficiency, and (c) absorption cross section in unreconstructed Si$_{66}$H$_{64}$ and reconstructed Si$_{66}$H$_{40}$ Si QDs.  Excitation energies are shown with respect to the energy of lowest excitonic state, $E^{X}_{|1>}$, in corresponding QDs. Solid curves in (a) are obtained in the same way as described in Fig. 2 (a).

\item[Figure \ 4]
Exciton transport (a) rate and (b) efficiency in 1.1 nm Si$_{35}$ QDs with different surface termination. Five surface passivants were considered: H, CH$_{3}$, F, Cl, and OH . In (a), results obtained from $1/R^{6}$ scaling of FT are shown as solid curves while dashed curves show $1/R^{14}$ fit to the transport rates from CH$_{3}$, F, Cl, and OH passivants.
\end{description}
\newpage

% Add figures
\begin{figure}
 \includegraphics[scale=0.75]{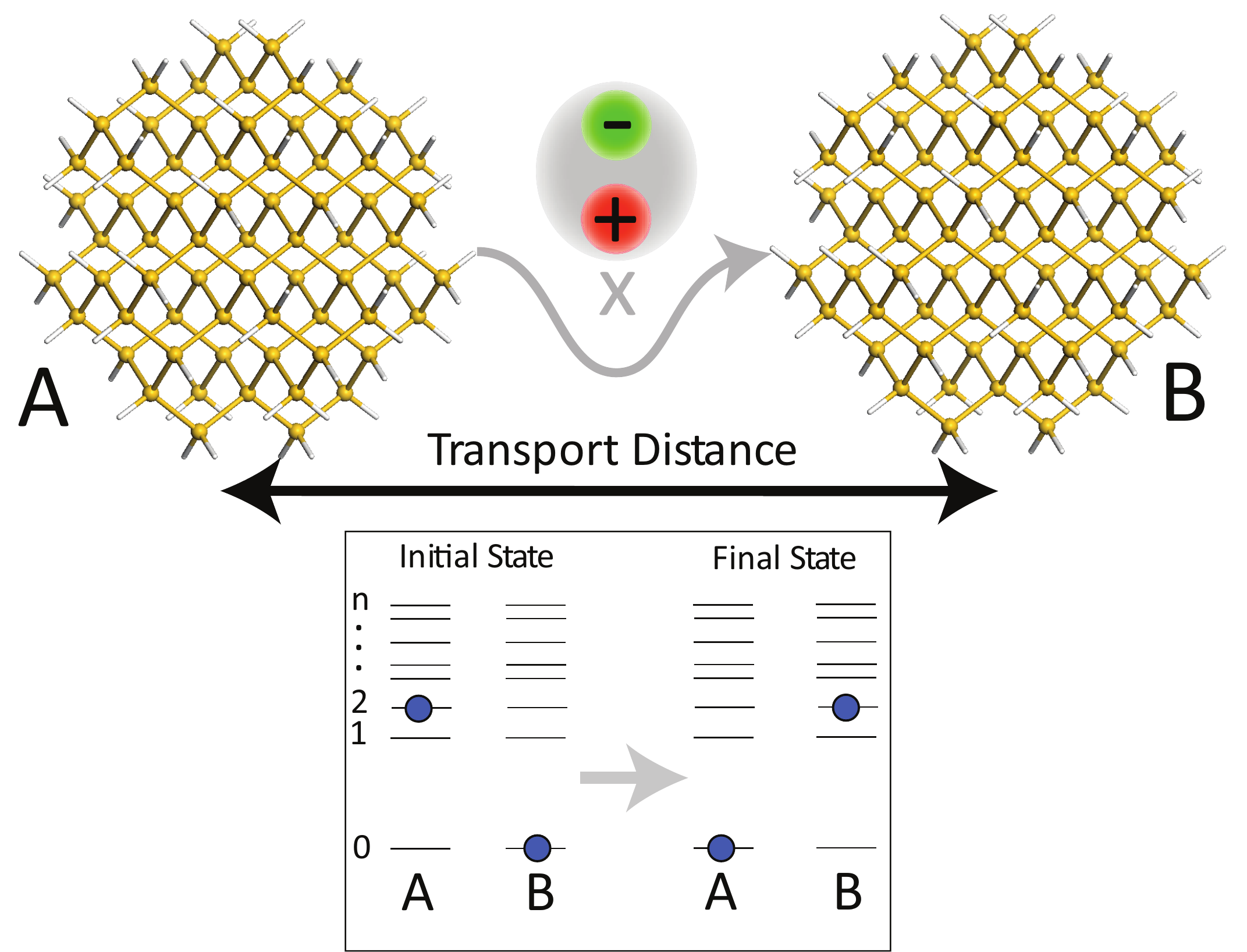}
\newline
\begin{center} 
\Large Figure 1 
\label{fig1}
 \end{center}
\end{figure}
\newpage

\begin{figure}[ptb]\begin{center}$
\begin{array}{cc}
\includegraphics[width=0.75\textwidth]{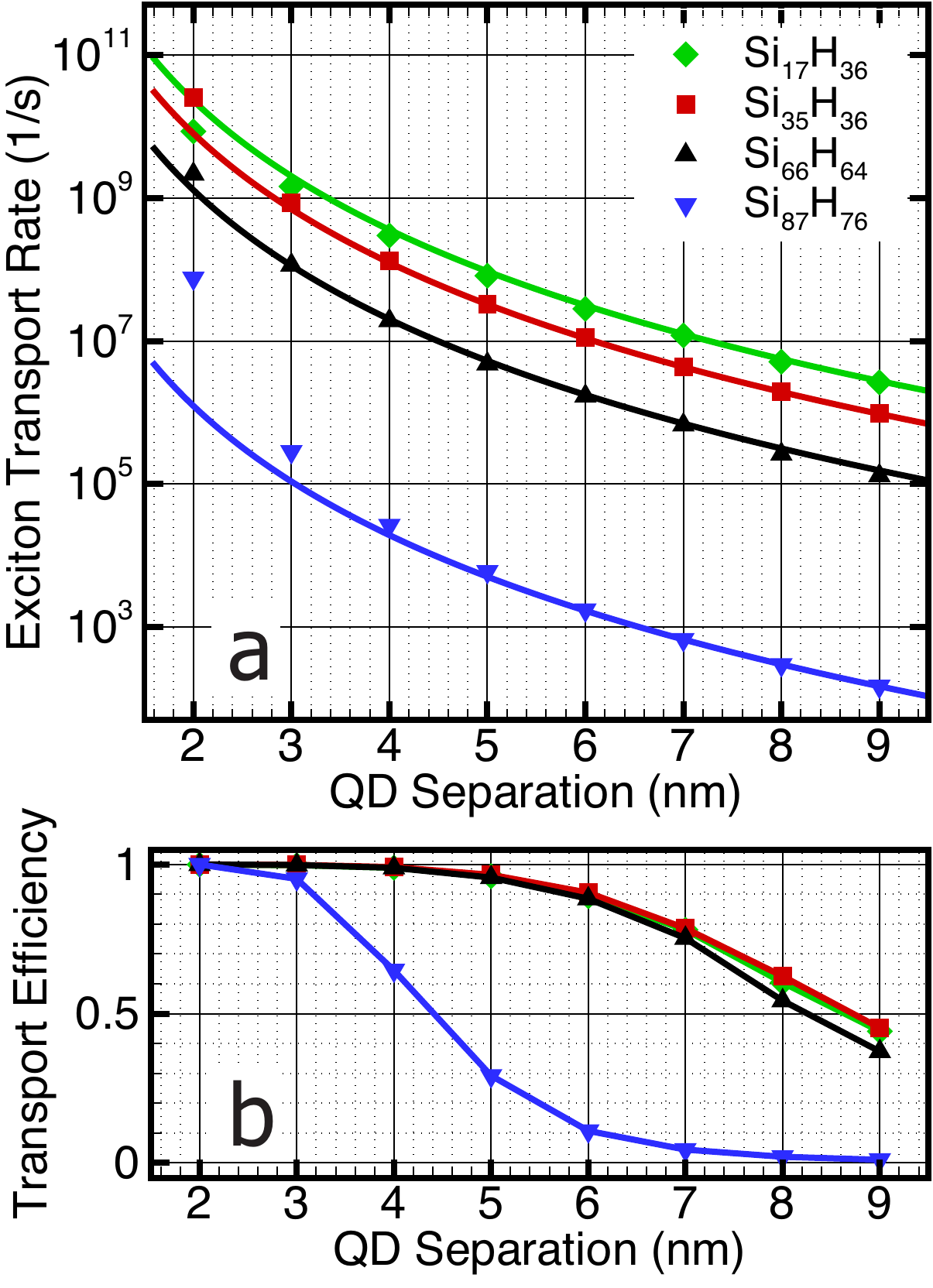} 
\end{array}$
\end{center}
\begin{center}
\Large Figure 2
\label{fig2}
\end{center}\end{figure}
\newpage

\begin{figure}[ptb]\begin{center}$
\begin{array}{cc}
\includegraphics[width=.6\textwidth]{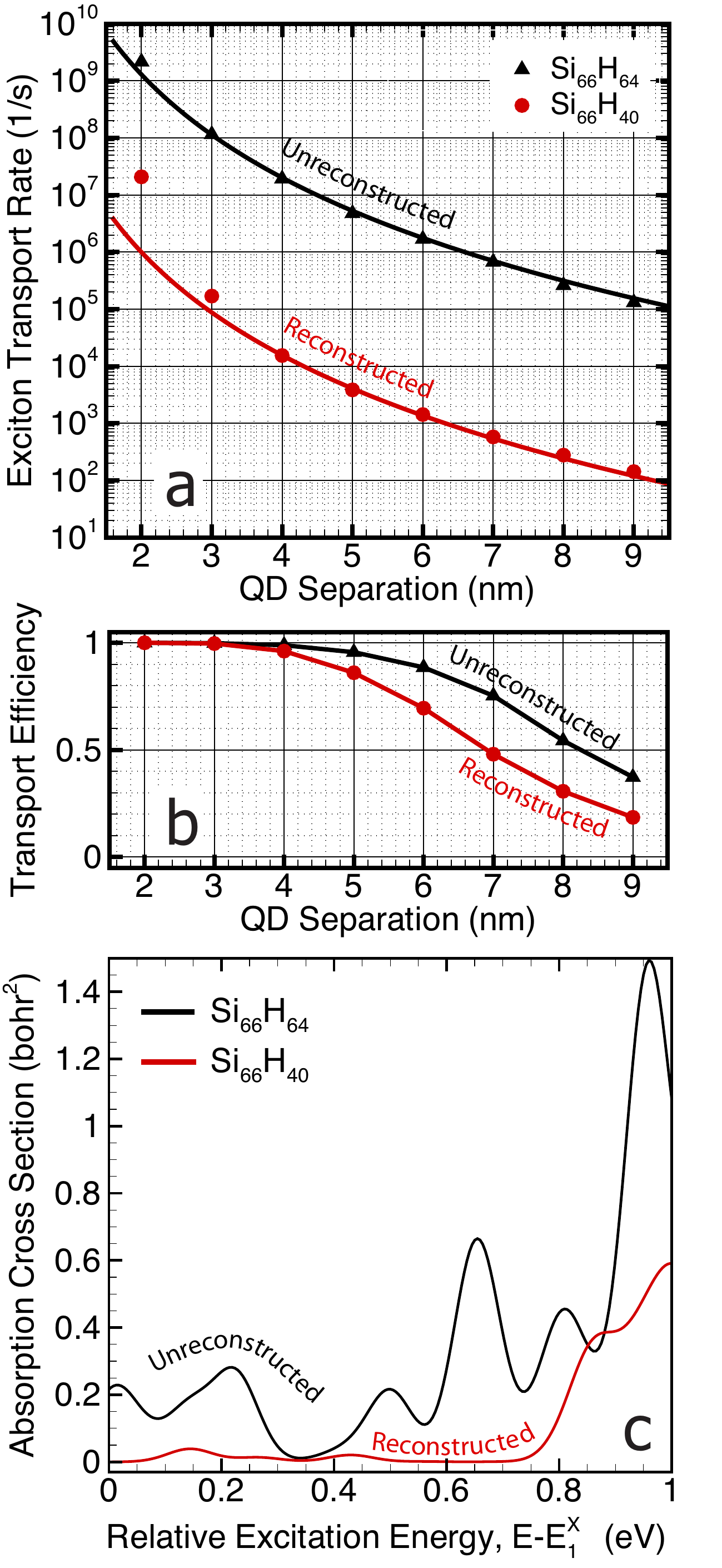} 
\end{array}$
\end{center}
\begin{center}
\Large Figure 3
\label{fig3}
\end{center}\end{figure}
\newpage

\begin{figure}[ptb]\begin{center}$
\begin{array}{cc}
\includegraphics[width=.75\textwidth]{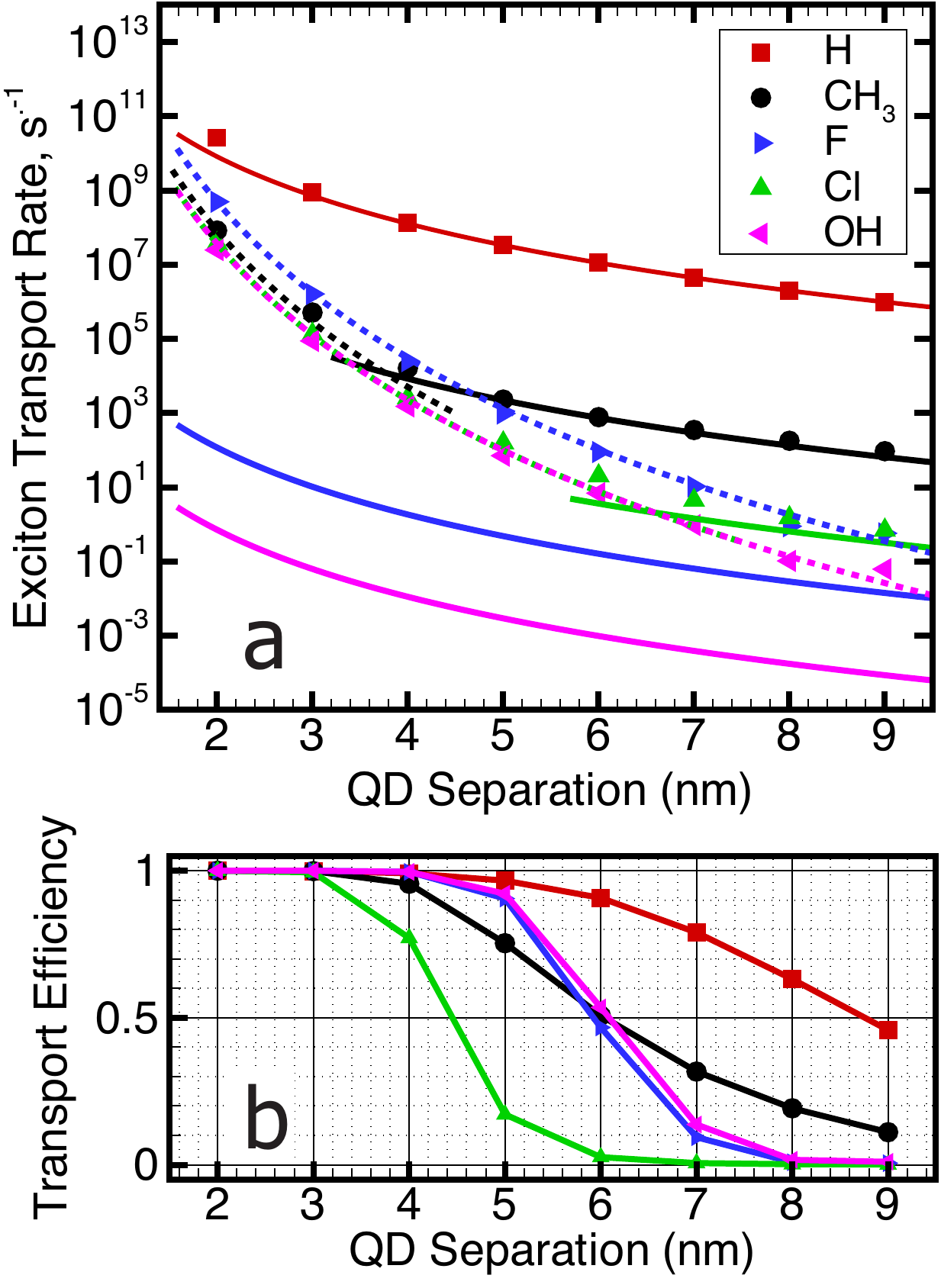} 
\end{array}$
\end{center}
\begin{center}
\Large Figure 4
\label{fig4}
\end{center}\end{figure}
\newpage

\end{document}

\section{Supporting Information}
To understand the influence of surface coverage of OH ligands on the exciton transport in Si QDs, we calculate the exciton transport rate and efficiency for Si$_{66}$ QDs with 25\% and 62.5\% surface coverage of OH ligands and compare them together with the one (Si$_{66}$H$_{64}$) without any OH passivant. In the partial OH passivation, hydrogen atoms are substituted with OH ligands to create partial coverage of OH on the QD surface. The structures are then fully relaxed within LDA. Note that full coverage of OH ligands on Si$_{66}$ is hindered by the steric effects of the passivants on the surface. The impact of replacing hydrogen passivation to OH on exciton transport can be clearly seen in Fig. S1. The delocalization of electron wavefunctions to the surface region due to more electronegative OH on the dot surface results in the vanishing dipole moments between the orbitals. As the OH coverage on the QD surface is increased, the exciton transport rate decreases for all QD separations.  The exciton transport efficiency (Fig. S1(b)) is nearly unity at QD separations less than 4 nm for all three OH coverages due to the much faster exciton transport rates as compared to the PL rates in the dots. The influence of surface coverage of OH ligands on the transport efficiency, however, becomes apparent at large dot separation as increasing OH coverage results in significant lower transport efficiency.

\section*{Figure captions}

\begin{description}
\item[Figure \ S1]
Exciton transport a) rate and b) efficiency in 1.3 nm Si$_{66}$ QDs with different surface coverage of OH ligands, namely 0\% (Si$_{66}$H$_{64}$), 25\%, and 62.5\% surface coverage.
\end{description}
% Add figures
\begin{figure}
 \includegraphics[width=.75\textwidth, keepaspectratio, clip]{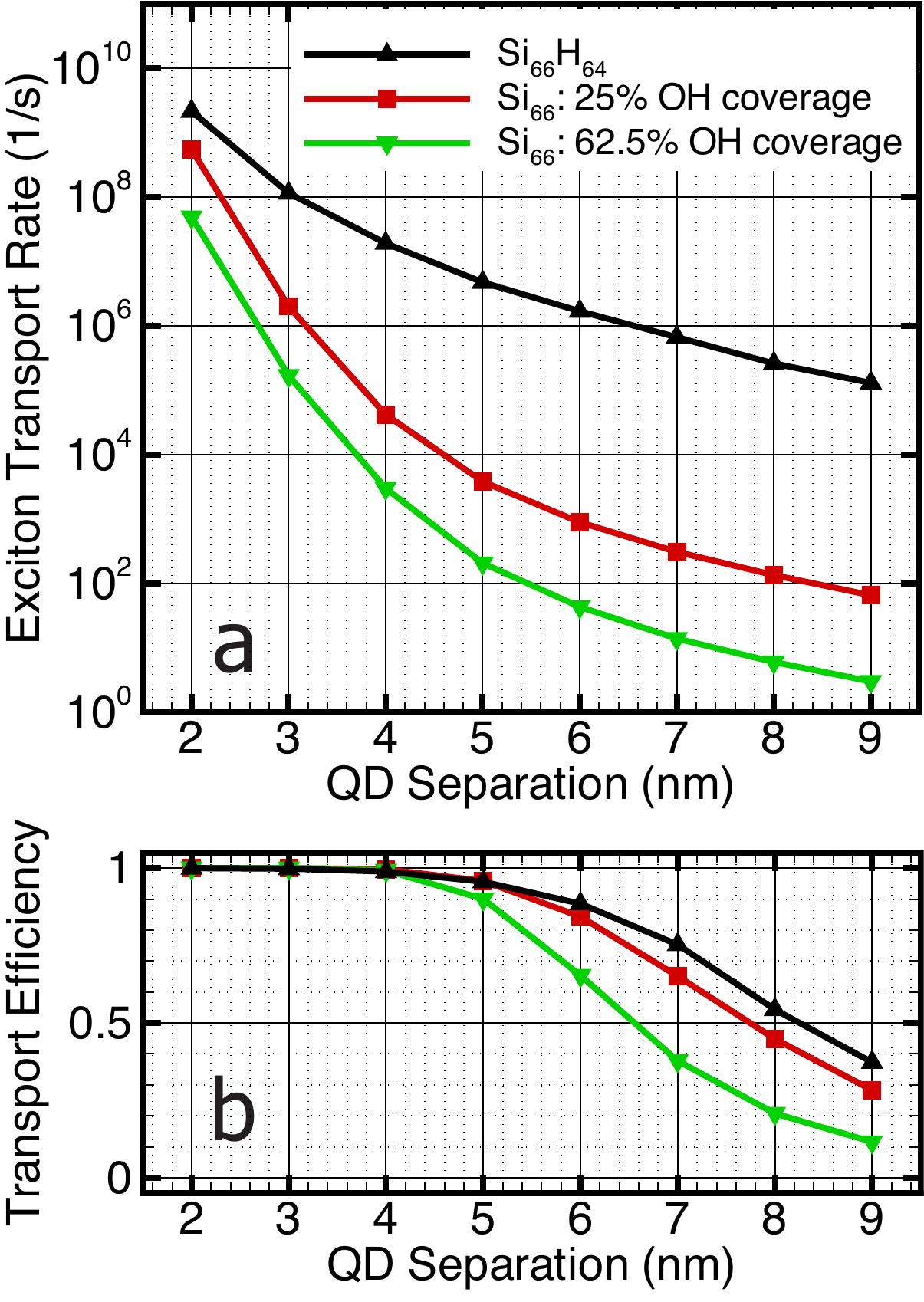}
\newline
\begin{center} 
\Large Figure S1 
 \end{center}
\end{figure}